\documentclass[aps,pra, superscriptaddress, floatfix, twocolumn]{revtex4} 
\newcounter{qcounter}

\usepackage{graphicx}
\usepackage{graphics}
\usepackage{amssymb}
\usepackage{amsmath}
\usepackage{bm}
\usepackage{epsfig}
\usepackage{latexsym}
\usepackage{color}
\usepackage{subfig}

\newcommand{\ket}[1]{| #1\rangle}
\newcommand{\bra}[1]{\langle #1 |}

\def\ketc[#1]{\vert #1 \rangle}
\def\brac[#1]{\langle #1 \vert}

\newcommand{\expect}[1]{\langle{#1}\rangle}

\newcommand{\beq}{\begin{equation}}
\newcommand{\eeq}{\end{equation}}
\newcommand{\bqa}{\begin{eqnarray}}
\newcommand{\eqa}{\end{eqnarray}}
\newcommand{\nn}{\nonumber}

\newcommand{\erf}[1]{Eq.~(\ref{#1})}
\newcommand{\dg}{^\dagger}

\newcommand{\BQIC}{Berkeley Quantum Information and Computation Center, Berkeley, California 94720 USA}
\newcommand{\DeptChem}{Department of Chemistry, University of California, Berkeley, California 94720 USA}
\newcommand{\NTU}{Department of Chemistry, National Taiwan University, Taipei City 106, Taiwan}

\begin{document}

\title{Environmental correlation effects on excitation energy transfer in photosynthetic light harvesting}

\author{Mohan Sarovar}
\email{msarovar@berkeley.edu}
\affiliation{\BQIC} 	\affiliation{\DeptChem} 

\author{Yuan-Chung Cheng}
\affiliation{\NTU}
 
 \author{K. Birgitta Whaley}
\affiliation{\BQIC}		\affiliation{\DeptChem} 

\begin{abstract}
Several recent studies of energy transfer in photosynthetic light harvesting complexes have revealed a subtle interplay between coherent and decoherent dynamic contributions to the overall transfer efficiency in these open quantum systems. In this work we systematically investigate the impact of temporal and spatial correlations in environmental fluctuations on excitation transport in the Fenna-Matthews-Olson photosynthetic complex. We demonstrate that the exact nature of the correlations can have a large impact on the efficiency of light harvesting. In particular, we find that (i) spatial correlations can enhance coherences in the site basis while at the same time slowing transport, and (ii) the overall efficiency of transport is optimized at a finite temporal correlation that produces maximum overlap between the environmental power spectrum and the excitonic energy differences, which in turn results in enhanced driving of transitions between excitonic states.
\end{abstract}

\maketitle

\section{Introduction}
Recent experimental revelations of long-lived electronic coherence in photosynthetic light harvesting systems \cite{Eng.Cal.etal-2007, Lee.Che.etal-2007, Col.Won.etal-2010, Pan.Hay.etal-2010} and conjugated polymers \cite{Col.Sch-2009} have prompted a renewed examination of energy transport in densely packed molecular aggregates. In particular, the effect of quantum coherent dynamics, and its interplay with environmental decoherence and dissipation have been closely scrutinized lately (e.g. \cite{Gaa.Bar-2004, Reb.Moh.etal-2009,Moh.Reb.etal-2008,Reb.Moh.etal-2009a, Ple.Hue-2008, Car.Chi.etal-2009, Ish.Fle-2009a}). Combined with earlier detailed modeling of excitation transport in photosynthetic light harvesting (e.g. \cite{Ren.May.etal-2001,Sch.Fle-2005,Che.Fle-2009}), a preliminary understanding of the complex dynamics of excitation transport is being molded. However, the picture is far from complete. Most photosynthetic light harvesting complexes (LHCs) are surrounded by protein structures that serve multiple functions, including maintenance of structural stability \cite{Cog.Gal.etal-2006} and creation of energy landscapes that facilitate energy transfer to reaction centers in the core chromophoric networks of LHCs \cite{Muh.Mad.etal-2007}. These same protein structures also provide a dynamic environment that interacts with the chromophore molecules that carry the excitation energy. The dynamics of this environment are complex and not very well characterized. Environmental fluctuations are generally correlated in time \cite{Ren.May.etal-2001,Ren.Mar-2002,Fle.Cho-1996} and are also believed to be spatially correlated \cite{Eng.Cal.etal-2007, Lee.Che.etal-2007}. The effects of these correlations on photosynthetic energy transport are generally not well understood (although recent experimental results suggest that spatial correlations may be directly responsible for long-lived electronic coherence \cite{Lee.Che.etal-2007, Col.Sch-2009}). Further, new dynamical models capable of simulating some of these correlations (e.g. \cite{Ish.Fle-2009, Naz-2009}) suggest that their effects are quite significant.  Thus it is important to determine the range of possible consequences of temporal and spatial correlations of environmental fluctuations on excitation energy transfer both within a single LHC and between networked photosynthetic units.

In this paper we undertake a numerical study of these effects, analyzing how the spatiotemporal correlations of environmental fluctuations affect energy transport in a single LHC. We use a specific system, the Fenna-Matthews-Olson (FMO) bacteriochlorophyll complex \cite{Fen.Mat-1975, Cam.Bla.etal-2003} as a prototypical model for our study because its structural and energetic properties are particularly well characterized \cite{Li.Zho.etal-1997,Cam.Bla.etal-2003}. We examine the effects of environmental fluctuations in a systematic fashion, employing model forms of correlations in time and space in order to assess the generic effects of these correlations on energy transfer efficiency. The results obtained here with FMO can thereby be taken as indicative of the generic effects of environmental fluctuations on other light harvesting complexes.

The FMO complex is a small protein in green sulfur bacterium that acts as a highly efficient energy transfer wire connecting chlorosomes, i.e., light collecting pigment arrays, to photosynthetic reaction centers.  Structurally, the FMO protein is a trimer whose monomers are believed to function independently \cite{Ado.Ren-2006}. Each monomer contains seven bacteriochlorophyll-\textit{a} (BChla) molecules embedded within a protein scaffold. Recent studies have determined the orientation of the FMO complex within the inter-membrane region between the chlorosome antenna and reaction center \cite{Ado.Ren-2006, Wen.Zha.etal-2009}. They present strong evidence that the reaction center is strongly coupled to BChla 3 and that the excitation energy enters an FMO monomer from the chlorosomes via BChla 1 or BChla 6. Below, we shall refer to the individual BChla molecules as \textit{sites} or \textit{chromophores}.

Under moderate laser driving (or \textit{in vivo}) there is at most one excitation in a single FMO complex. In these conditions, the reversible dynamics of the electronic degrees of freedom are described by the so-called Frenkel exciton Hamiltonian: $H_{\textrm{el}} = \sum_{j=1}^7 E_j \ket{j}\bra{j} + \sum_{j=1}^7 \sum_{i>j}^7 J_{ij}(\ket{j}\bra{i} + \ket{i}\bra{j})$. Here $\ket{j}$ represents the state where only the $j$th chromophore is excited and all other chromophores are in their electronic ground states. $E_j$ is the transition energy of chromophore $j$, including any static shifts due to the interactions with the protein environment, and $J_{ij}$ describe the excitonic coupling between chromophores $i$ and $j$. We adopt a classical stochastic model of the FMO environment and describe the interactions between excitations and surrounding protein environments as fluctuations of the chromophore transition energies: $H_{\textrm{s}} = \sum_{j=1}^7 \Delta_j (t) \ket{j}\bra{j}$, where $\Delta_j(t)$ are time dependent random variables whose properties we shall describe shortly. We ignore fluctuations in the off-diagonal couplings for simplicity, but our treatment below can be generalized to treat such noise as well. This model is equivalent to a stochastic Liouville treatment of excitation dynamics and will lead to dephasing in the site basis when the dynamics are averaged over the random process \cite{Hak.Str-1973, Sil-1976, Rip-1993}.
Since we are treating the fluctuating phonon environment as a classical quantity, dynamics under this stochastic Liouville treatment are only exact for an infinite temperature phonon environment, although they are known to become increasingly accurate for high temperature baths \cite{Rip-1993}. In the context of this work, the primary utility of this model comes from the fact that it allows one to numerically incorporate any spatial and temporal correlation of the environment into the dynamics of the chromophores, and thus enables systematic studies of a range of correlated bath dynamics. This is difficult or even impossible for alternative treatments of the chromophore-environment interactions such as Redfield equations, Markovian master equations, or cumulant expansion techniques \cite{Nit-2006}. Since the main conclusions we draw from the present study will pertain to the \textit{relative} effects of environmental correlations, this stochastic model will be sufficient for our purposes. In summary, the dynamics of the FMO complex will be described by the following master equation:
\beq
\frac{\textrm{d}\rho}{\textrm{d}t} = \frac{-i}{\hbar} [H, \rho]  + \gamma_l \sum_{j=1}^7 \mathcal{D}[\sigma_j^-]\rho + \gamma_t \mathcal{D}[\sigma_{\textrm{trap}}^{+}\sigma_3^-]\rho
\label{eq:me}
\eeq
where $H\equiv H_{\textrm{el}} + H_{\textrm{s}}$, and the second and third terms describe, respectively, radiative excitation decay and excitation trapping at site 3 due to interaction with the reaction center. The Lindblad superoperator $\mathcal{D}[A]\rho \equiv A\rho A\dg -\frac{1}{2} A\dg A \rho - \frac{1}{2} \rho A\dg A$ for any operator $A$, and $\sigma_j^- = \ket{0}\bra{j}$ is a lowering operator ($\ket{0}$ denotes the electronic ground state of all seven chromophores). 

The stochastic noise processes $\Delta_{j}(t)$ are Gaussian with zero mean, a reasonable form for noise terms resulting from coupling to a dynamic protein environment \cite{Ren.May.etal-2001}. The temporal correlations are taken to be of exponential form: $\expect{\Delta_{j}(t)\Delta_{j}(t+\tau)} = \Delta_{0}^{2}\exp(-|\tau |/\tau_{c})$, where the correlation time, $\tau_{c}$, is a free parameter which we will vary in the simulations below. The magnitude of the noise variation, $\Delta_{0}$, characterizes the size of the environmental fluctuations, which we expect to be a function of the environmental dynamics and temperature. Consistent with the stochastic Liouville treatment, we use a high temperature approximation for the variance of phononic fluctuations (e.g. see Ch. 8 of Ref. \cite{Muk-1999}) to arrive at $\Delta_{0}^{2} = 2E_{R} k_{B} T$, where $E_{R}$ is the reorganization energy and $T$ is the temperature. The true form of the decay of temporal correlations in protein dynamics is complex. It is typically a combination of Gaussian decay at initial times and multiple exponential decays at longer times. Due to the complexity of simulating such a correlation and the lack of detailed knowledge of temporal correlation in the FMO environment, we have chosen here to simulate the temporal correlation decay with a single exponential. At the temperatures and timescales relevant to light harvesting complex dynamics this approximation is reasonable and corresponds to a coarse-graining of environmental dynamics \cite{Ish.Fle-2009}. This is also the justification provided for the use of the overdamped Brownian oscillator model commonly employed to model protein dynamics in LHCs (e.g. \cite{Ren.May.etal-2001, Bel.Cur.etal-2009, Ish.Fle-2009b}). In fact, the single exponential decay form we use is the high-temperature limit of the symmetrized correlation function for the overdamped Brownian oscillator model -- where the Matsubara terms can be ignored \cite{Ish.Fle-2009b}. Another justification for this form of temporal correlation decay comes from Doob's theorem \cite{vanK-2007}, which states that any stationary, Gaussian and Markov random process (all plausible properties for the environmental fluctuations in pigment-protein complexes at physiological temperatures) must possess an exponentially decaying temporal correlation. 

Spatial correlations of the noise processes are described by a matrix $\mathbf{C}$ with elements $\mathbf{C}_{ij} = \frac{1}{\Delta^{2}_{0}}\expect{\Delta_{i}(t)\Delta_{j}(t)}$. The spatial correlation of protein dynamics in photosynthetic complexes is not a well studied subject and there are currently no precise characterizations of either the nature or the extent of spatial correlation between electron-phonon couplings or phonon fluctuations. In this study we will therefore explore the effects of generic forms for $\mathbf{C}$, within the restriction of having positive correlations between fluctuations at different chromophores (i.e., positive definite matrix $\mathbf{C}_{ij}\geq 0$).  Specifically, we will examine the following three models:

\begin{figure*}[t!]
\centering 
\subfloat[~Dimerized correlations] { 
\label{fig:i1_77:a} 
\includegraphics[width=7.5cm]{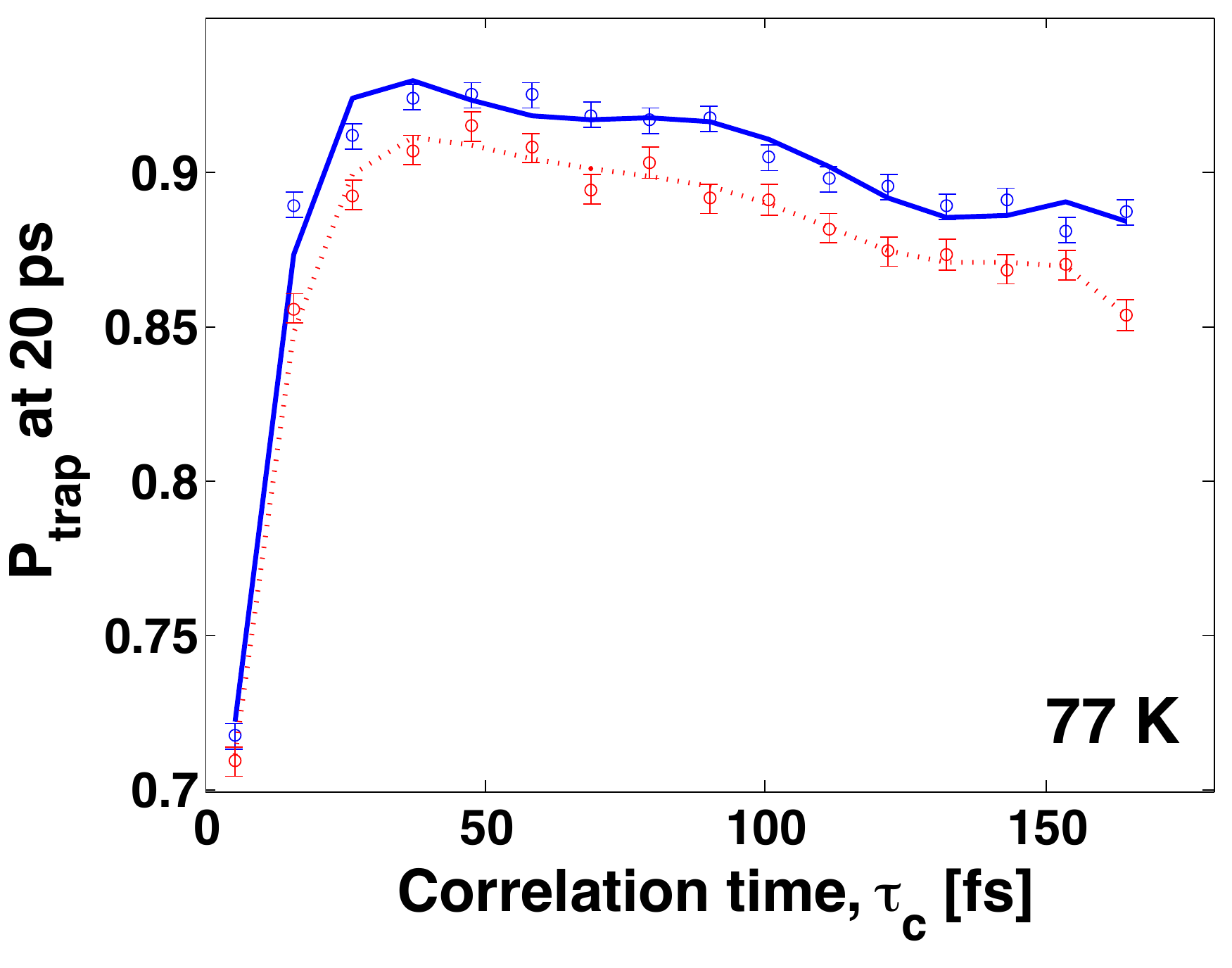}
} 
\hspace{1cm}
\subfloat[~Exponential correlations] { 
\label{fig:i1_77:b}
\includegraphics[width=7.5cm]{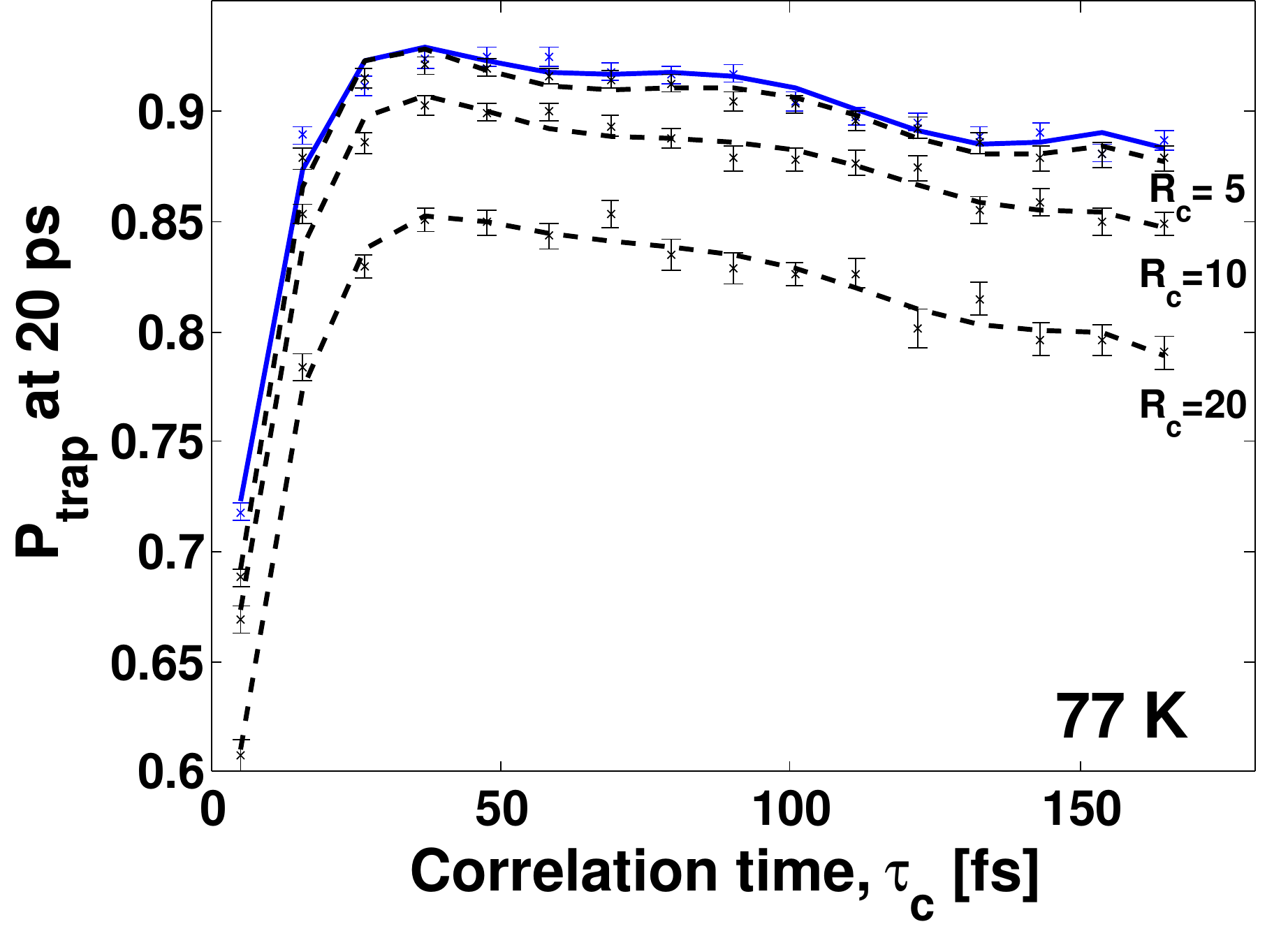}
}
\caption{(Color online) Trapping probability at 20 ps as a function of the temporal correlation of the FMO environment, for various spatial correlation models.   The initial state is an excitation localized on BChla 1, and the temperature in both panels is T=77K. The curves are polynomial fits to the data points that indicate the general trend, and the error bars show the standard deviation of the average taken over 100 sample evolutions. In both panels, the blue (solid) curve corresponds the case of no spatial correlation $\mathbf{C}^{N}$, the red (dotted) curve corresponds to the case of dimerized spatial correlation $\mathbf{C}^{D}$, and the black (dashed) curves correspond to exponential spatial correlation $\mathbf{C}^{E}$, with the labeled correlation radii. Panel (a) compares the trapping probability for the case of no spatial correlation to the case of dimerized correlations, and panel (b) compares no correlation with exponential correlations.} 
\label{fig:i1_77}
\end{figure*}

\begin{figure*}[t!]
\centering 
\subfloat[~Dimerized correlations] { 
\includegraphics[width=7.5cm]{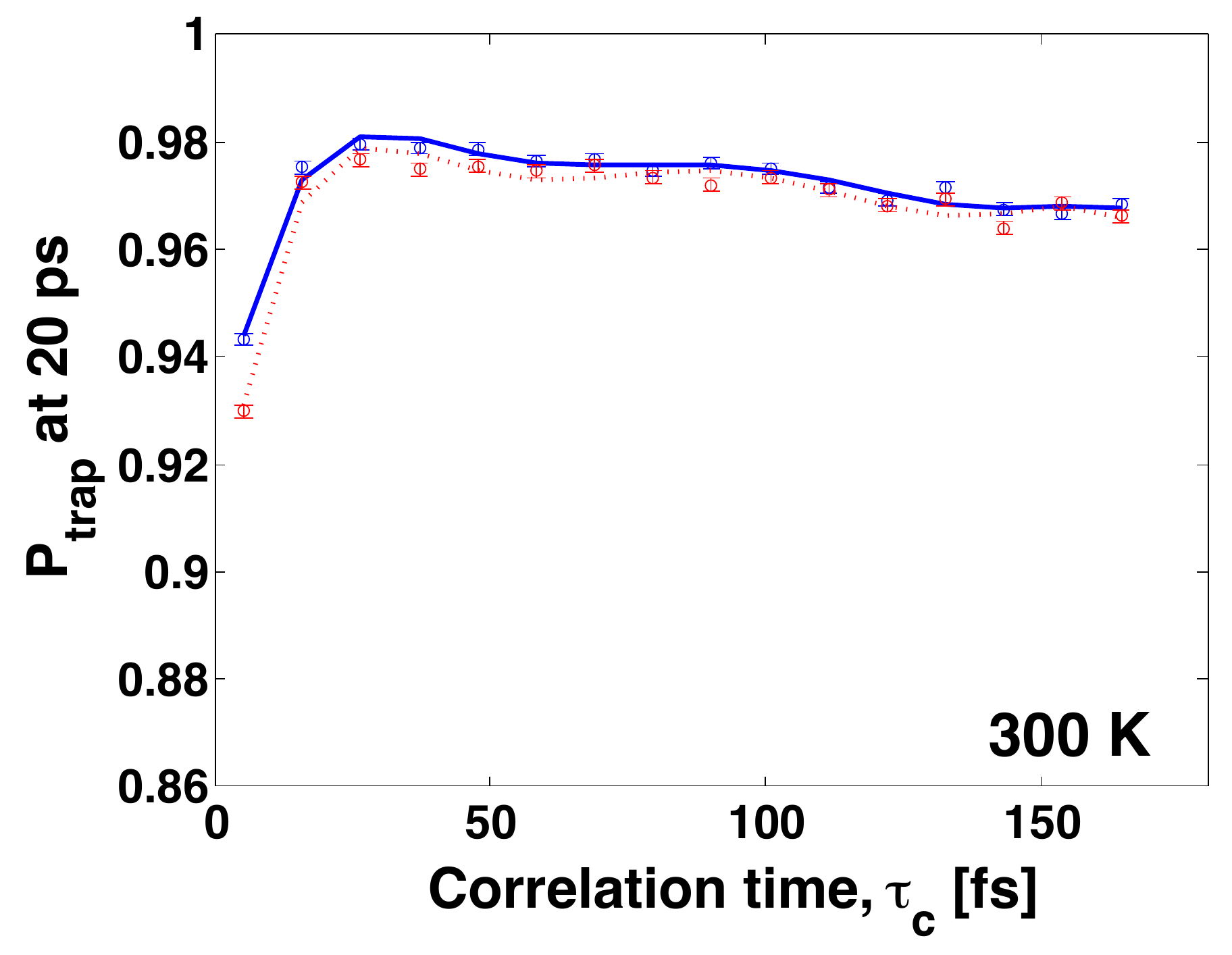}
} 
\hspace{1cm}
\subfloat[~Exponential correlations] { 
\includegraphics[width=7.5cm]{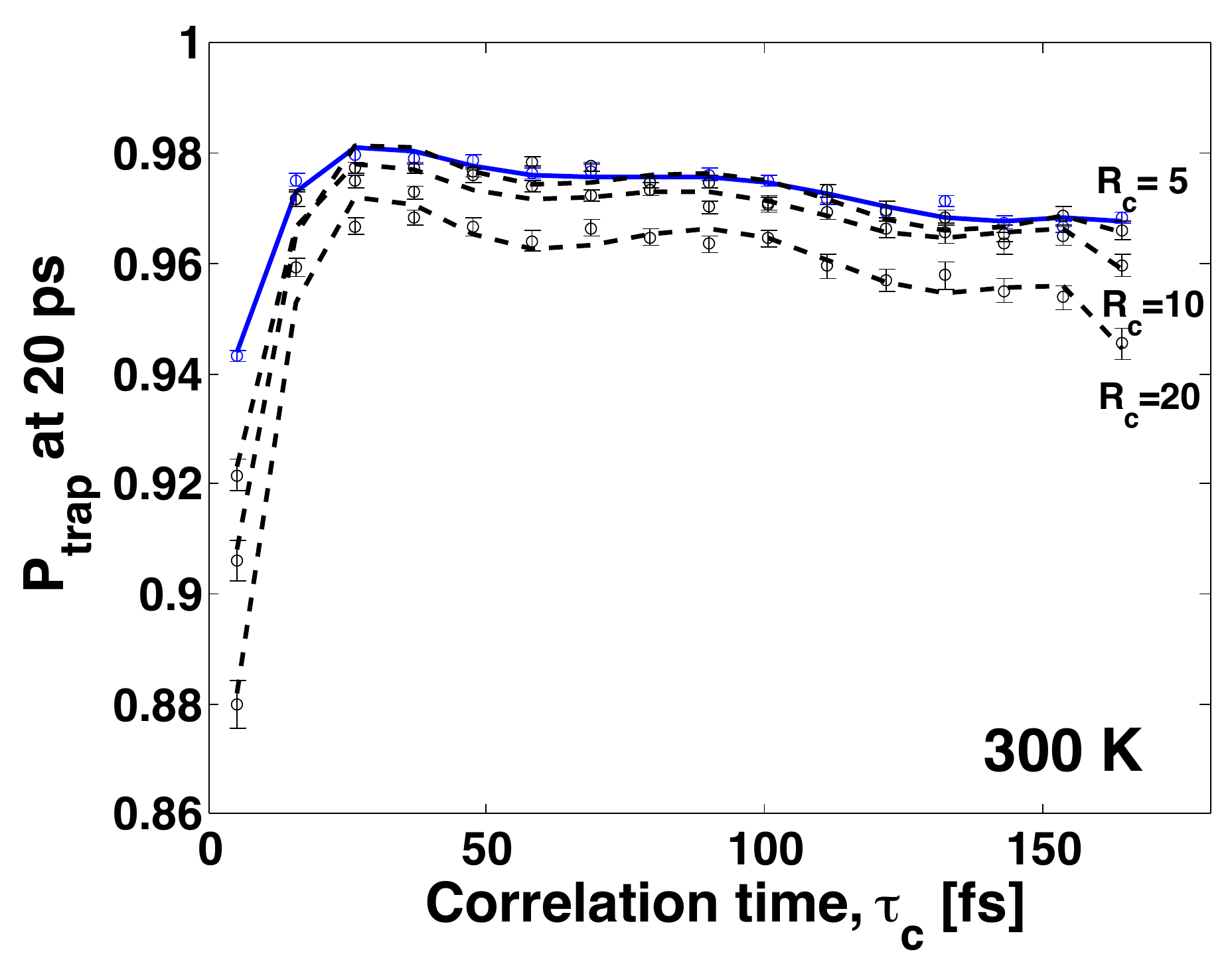}
}
\caption{(Color online) Trapping probability at 20 ps as a function of the temporal correlation of the FMO environment, for various spatial correlation models.  The initial state is an excitation localized on BChla 1, and the temperature in both panels is T=300K. The curves are polynomial fits to the data points that indicate the general trend, and the error bars show the standard deviation of the average taken over 100 sample evolutions. The color coding for the curves is the same as in Fig. \ref{fig:i1_77}. Panel (a) compares the trapping probability for the case of no spatial correlation to the case of dimerized correlations, and panel (b) compares no correlation with exponential correlations.} 
\label{fig:i1_300}
\end{figure*}

\begin{list}{\arabic{qcounter}.}{\leftmargin=1em}
\usecounter{qcounter}
\addtolength{\itemsep}{ -0.6\baselineskip}
\item \emph{No spatial correlations}. $\mathbf{C}_{ij}^{N} = \delta_{ij}$.
\item \emph{Dimerized correlations}. Excitons in FMO are mostly delocalized on two chromophores \cite{Bri.Ste.etal-2005}. The chromophores that are the most strongly coupled are the pairs: $1$-$2$, and $5$-$6$, and to a lesser extent, $4$-$5$ and $4$-$7$. This dimerization motivates us to use a correlation matrix, $\mathbf{C}^{D}$, with the only non-zero off-diagonal entries being: $\mathbf{C}_{12}^{D}=\mathbf{C}_{21}^{D}=\mathbf{C}_{56}^{D}=\mathbf{C}_{65}^{D}=0.9, \mathbf{C}_{45}^{D}=\mathbf{C}_{54}^{D}=\mathbf{C}_{47}^{D}=\mathbf{C}_{74}^{D}=0.4$. That is, only the strongly coupled chromophores experience correlated fluctuations.
\item \emph{Distance-dependent correlations}. The typical distances between chromophores in the \textit{C. tepidum} FMO monomer are known from its crystal structure \cite{Li.Zho.etal-1997,Cam.Bla.etal-2003}. It is common in the literature to assume spatial correlations that decay exponentially with distance: $\mathbf{C}^{E}_{ij} = e^{-d_{ij}/R_{c}}$, where $R_{c}$ is a variable correlation radius \cite{Ren.May-1998}.  However, microscopic justification for this form is an open question \cite{Ren-2009} and so we have carried out calculations with both exponential and inverse polynomial decay of spatial correlations. 
\end{list}
Taken together, these models 1 - 3 allow a systematic study of the effects of increasing spatial correlations.

\section{Results} 
\erf{eq:me} was numerically integrated using the Adams-Bashforth-Moulton multistep method. The temporal correlations in the noise processes were generated from white noise by passing the latter through a finite impulse response filter \cite{Des-2002}. At each time instant the spatial correlation between fluctuations at different sites was realized by the standard technique of using the Cholesky decomposition of $\mathbf{C}$, see e.g. \cite{Joh-1994}. The electronic Hamiltonian, $H_{\textrm{el}}$, was formed using site energies and coupling strengths for \textit{C. tepidum} FMO from tables 1 (column 3) and 4 (trimer column) of Ref. \cite{Ado.Ren-2006}. Finally, we used values of $E_{R}= 35\textrm{cm}^{-1}$, $\gamma_{l} = 1 ~\textrm{ns}^{-1}$ and $\gamma_{t} = 1 ~\textrm{ps}^{-1}$, all of which are consistent with the most detailed experimental and theoretical literature on FMO \cite{Cho.Vas.etal-2005, Ado.Ren-2006}. 

\begin{figure*}[t!]
\centering 
\subfloat[~Dimerized correlations] { 
\includegraphics[width=7.5cm]{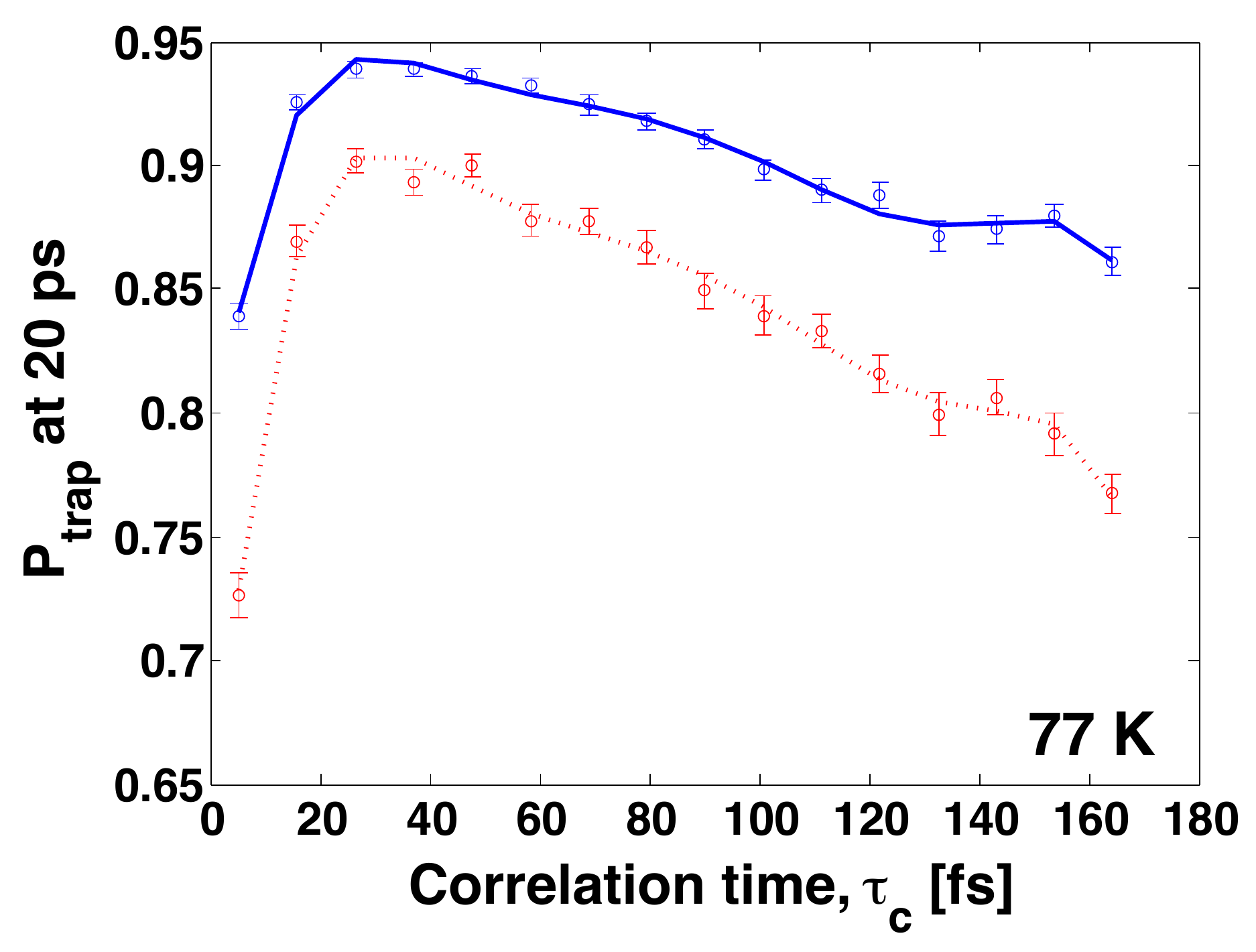}
} 
\hspace{1cm}
\subfloat[~Exponential correlations] { 
\includegraphics[width=7.5cm]{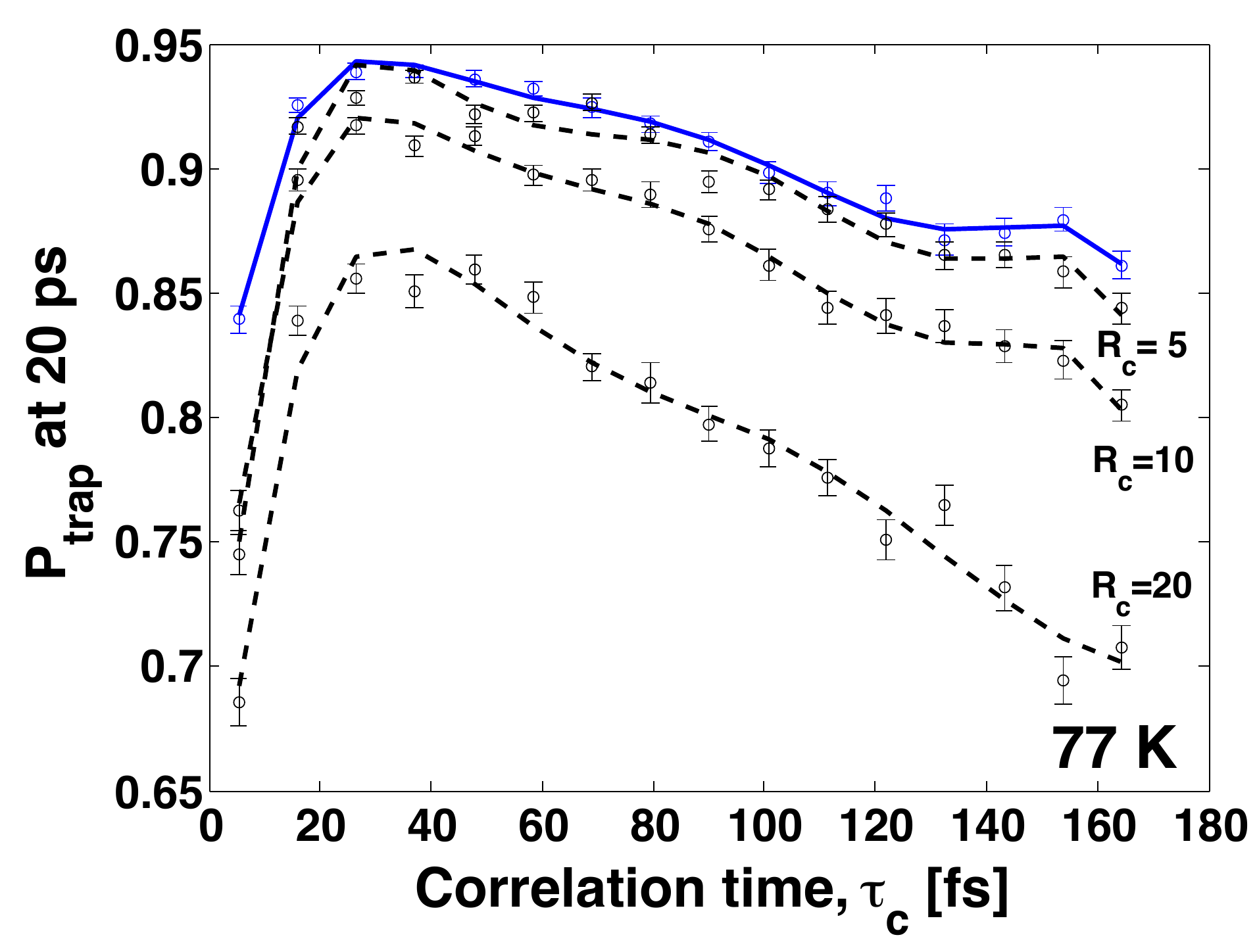}
}
\caption{(Color online) Trapping probability at 20 ps as a function of the temporal correlation of the FMO environment, for various spatial correlation models.  The initial state is an excitation localized on BChla 6, and the temperature in both panels is T=77K. The curves are polynomial fits to the data points that indicate the general trend, and the error bars show the standard deviation of the average taken over 100 sample evolutions. The color coding for the curves is the same as in Fig. \ref{fig:i1_77}. Panel (a) compares the trapping probability for the case of no spatial correlation to the case of dimerized correlations, and panel (b) compares no correlation with exponential correlations.} 
\label{fig:i6_77}
\end{figure*}

\begin{figure*}[t!]
\centering 
\subfloat[~Dimerized correlations] { 
\includegraphics[width=7.5cm]{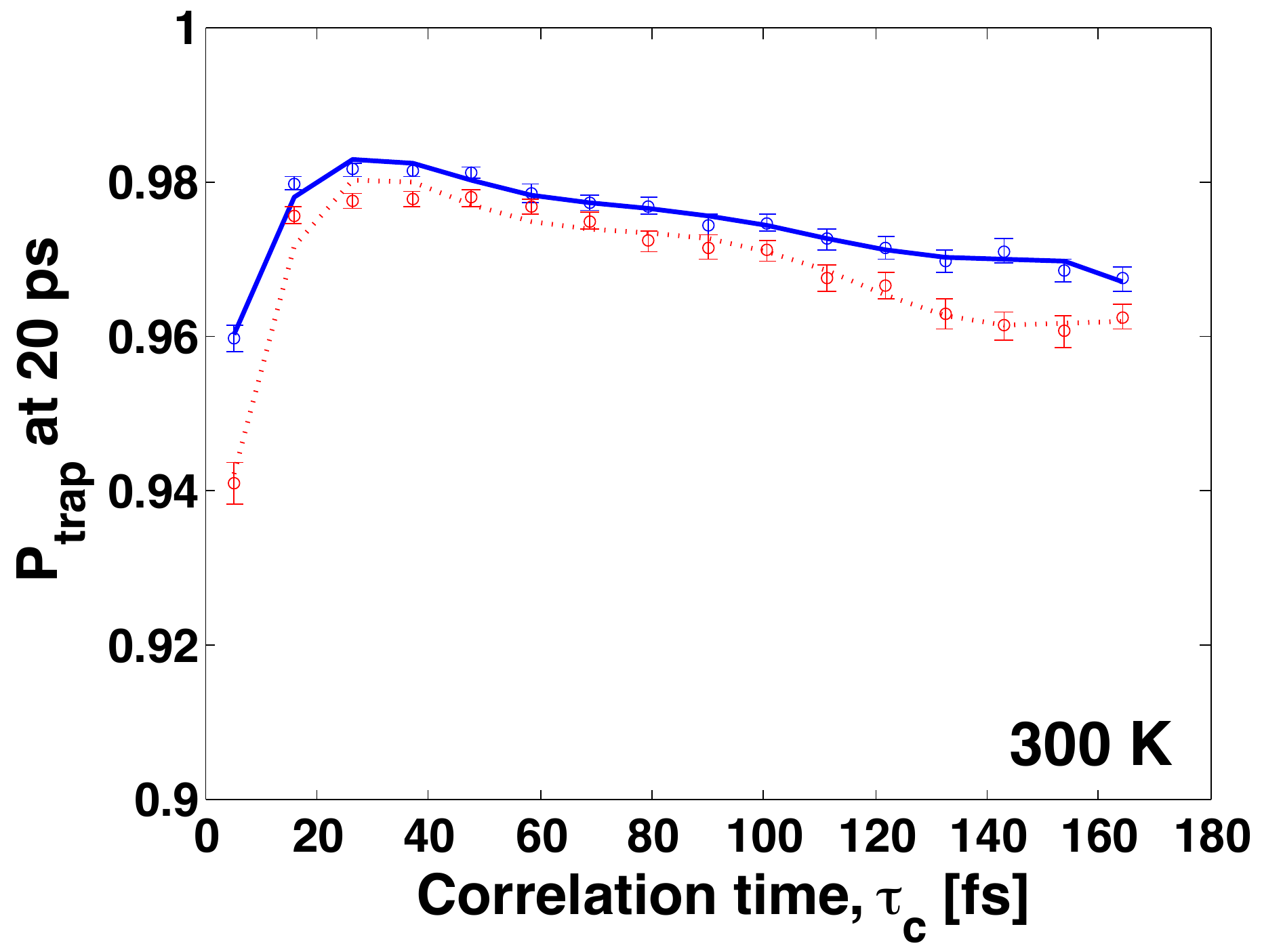}
} 
\hspace{1cm}
\subfloat[~Exponential correlations] { 
\includegraphics[width=7.5cm]{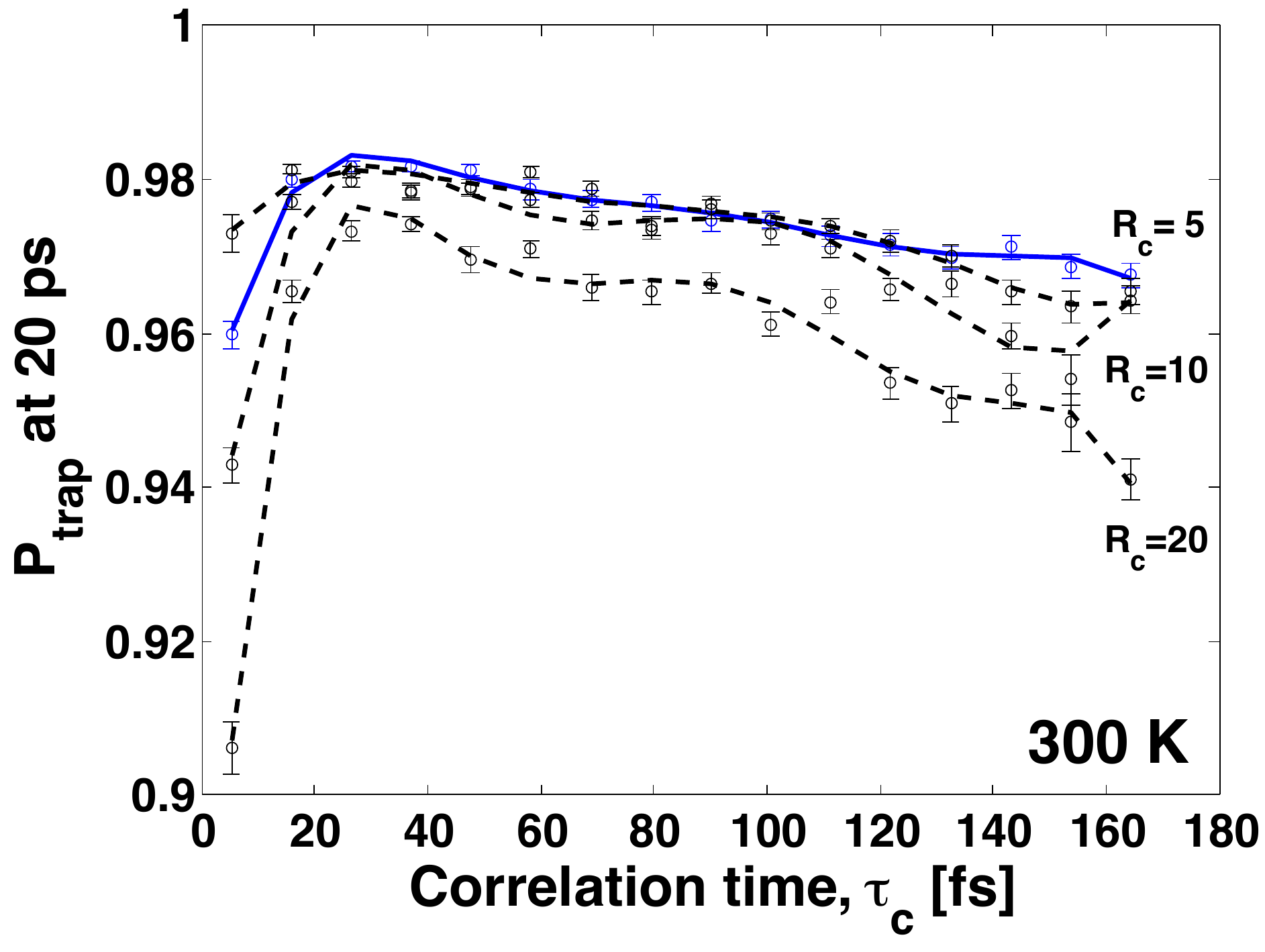}
}
\caption{(Color online) Trapping probability at 20 ps as a function of the temporal correlation of the FMO environment, for various spatial correlation models.  The initial state is an excitation localized on BChla 6, and the temperature in both panels is T=300K. The curves are polynomial fits to the data points that indicate the general trend, and the error bars show the standard deviation of the average taken over 100 sample evolutions. The color coding for the curves is the same as in Fig. \ref{fig:i1_77}. Panel (a) compares the trapping probability for the case of no spatial correlation to the case of dimerized correlations, and panel (b) compares no correlation with exponential correlations.} 
\label{fig:i6_300}
\end{figure*}

We focus primarily on the time evolution of the average trapping probability of the excitation. This is formally defined as: $P_{\textrm{trap}}(t) = \mathbb{E}_{\Delta}[\bra{\textrm{trap}}\rho(t)\ket{\textrm{trap}}]$, where $\rho(t)$ is the state of the system at time $t$, and $\mathbb{E}_{\Delta}[\cdot]$ denotes an ensemble average over instances of the stochastic process. In the simulations presented below, we perform this ensemble average over 100 instances. 
$P_{\textrm{trap}}$ provides a measure of the efficiency of excitation transport across the FMO protein, from the initial absorption at sites 1 or 6, to the trap, which represents the reaction center. Figures \ref{fig:i1_77} - \ref{fig:i6_300} show the average trapping probability at 20ps as a function of the temporal correlation time of the environmental fluctuations for various temperatures and initial states. In all the figures, the blue (solid) curve shows the trapping probability for the case of no spatial correlations, the red (dotted) curve shows $P_{\textrm{trap}}$ for the case of dimerized spatial correlations, and the black (dashed) curves shows the same quantity for exponential spatial correlations, with the labeled correlation radii. We choose to plot the average trapping probability at 20ps because we wish to compare relative efficiencies at times when transient effects are absent and thus are interested in the asymptotic behavior. However, we note that at earlier times the curves look qualitatively similar to those in Figs.~ \ref{fig:i1_77} - \ref{fig:i6_300} and the conclusions given below hold for all times in the interval $10-25$ps.

Several distinct features are evident from Figs. \ref{fig:i1_77} - \ref{fig:i6_300}. Firstly, with \emph{any} spatial correlation, there is an optimal temporal correlation time where the efficiency of excitation transport is maximized with respect to environmental fluctuations. For $T=77K$ this optimal time is $\sim 40$fs when the initial state is an excitation on site 1 (Fig. \ref{fig:i1_77}), and $\sim 30$fs when the initial state is an excitation on site 6 (Fig. \ref{fig:i6_77}). The efficiency dies off very quickly for shorter correlation times and more slowly for longer correlation times. A second feature evident in these figures is the large effect that spatial correlations have on the efficiency of excitation transport. Uncorrelated fluctuations provide the greatest efficiency and generally the efficiency decreases with increasing spatial correlation. There is very little difference between the uncorrelated spatial fluctuations case and exponential correlated case with the smallest correlation radius $R_{c}=5 \AA$, for both initial states and temperatures. And the dimerized correlations result in efficiencies that are similar to those produced by the exponential correlation with $R_{c}=10 \AA$. This is not surprising since the dimers in FMO are formed by pigments separated by roughly $10\AA$. Also, in comparing Figs. \ref{fig:i1_77} and \ref{fig:i6_77} we see that energy transfer from initial state 6 is more sensitive to spatiotemporal correlations than from initial state 1.

Comparison of Fig. \ref{fig:i1_77} with Fig. \ref{fig:i1_300} and Fig. \ref{fig:i6_77} with Fig. \ref{fig:i6_300} shows that the effect of increasing temperature, which in our model increases the variance of fluctuations, is to render the excitation transport less sensitive to temporal and spatial correlations. The optimal temporal correlation time is less pronounced; there is a wide plateau of comparable efficiencies across the range $30\textrm{fs}<\tau_{c}<90\textrm{fs}$ (Figs. \ref{fig:i1_300} and \ref{fig:i6_300}). Similarly, the variation of average trapping probability with spatial correlation is less at high temperature, although the general trend of decreased efficiency with increased spatial correlation still persists. These simulations also show that the average efficiency of transport is more robust to the initial state at higher temperatures -- i.e. there is less variation between Figs. \ref{fig:i1_300} and \ref{fig:i6_300} than between \ref{fig:i1_77} and \ref{fig:i6_77}. 

\begin{figure*}[t!]
\centering 
\subfloat[~$\tau_{c}=45\textrm{fs}$] { 
\includegraphics[width=7.5cm]{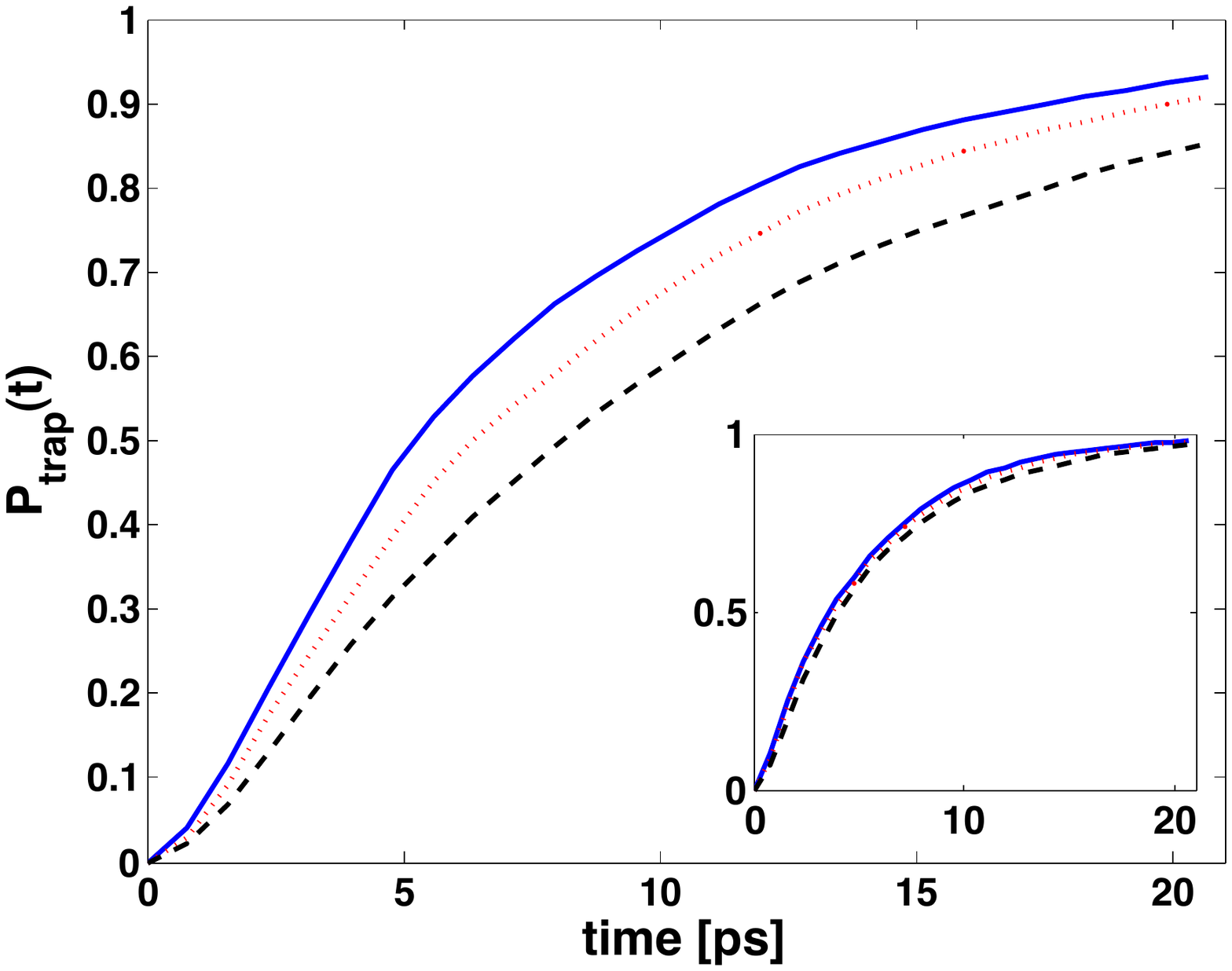}
} 
\hspace{1cm}
\subfloat[~$\tau_{c}=120\textrm{fs}$] { 
\includegraphics[width=7.5cm]{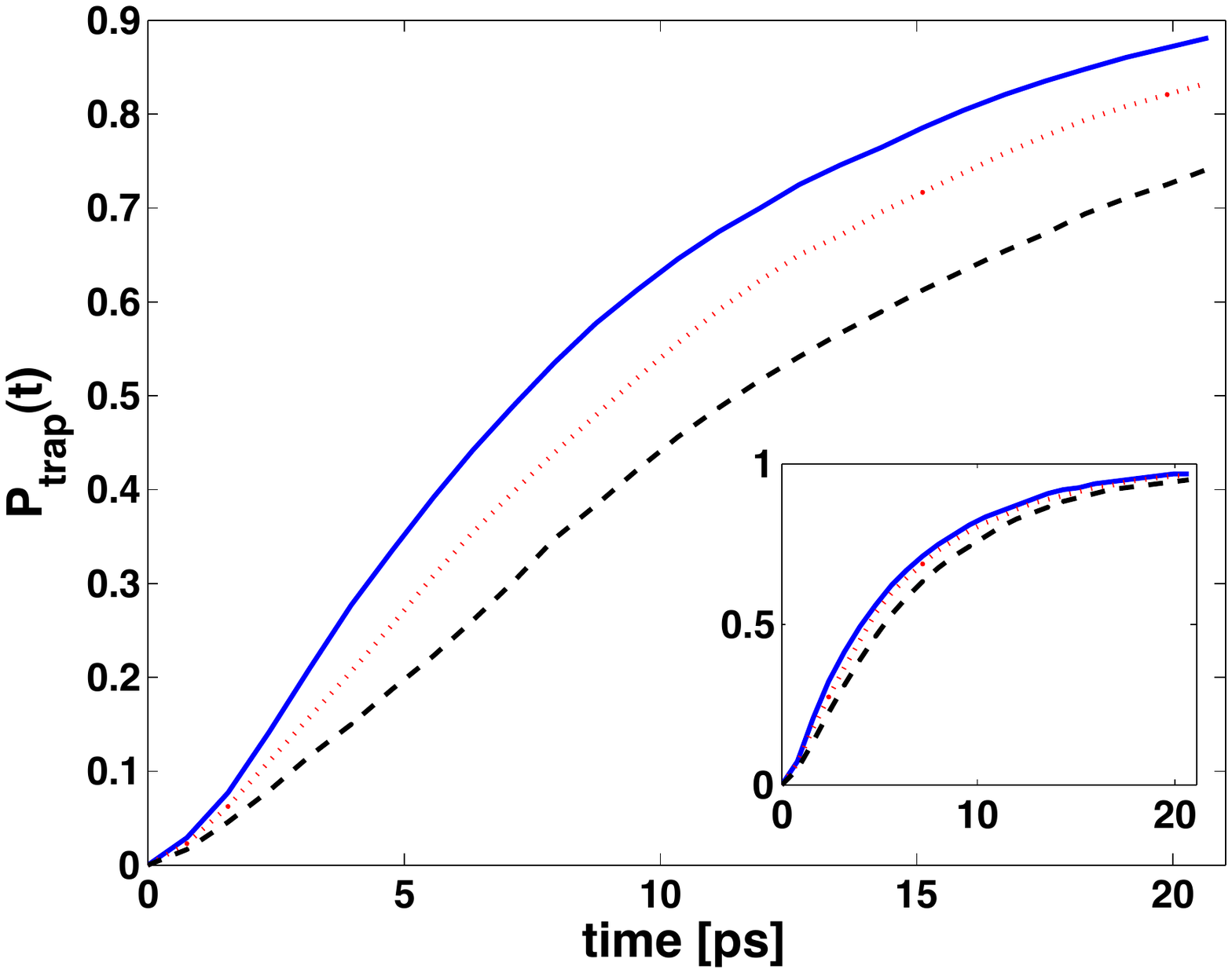}
}
\caption{(Color online) Average trapping probability as a function of time for calculations with two representative environmental correlation times, (a) $\tau_{c}=45\textrm{fs}$ and (b) $\tau_{c}=120\textrm{fs}$. In each case the main panel shows the behavior at $T=77K$ while the insets show behavior at $T=300K$.  The initial state is an excitation on BChla 6.  In both panels the blue (solid) curves are for no spatial correlations, $\mathbf{C}^{N}$, and the red (dotted) and black (dashed) curves are for exponential spatial correlations with $R_c=10 \AA$ and $R_c=20\AA$, respectively. The averages are taken over 100 sample evolutions. Error bars are omitted for clarity: the variation from these average curves is small.} 
\label{fig:ptrap}
\end{figure*}

\begin{figure}[t!]
	\includegraphics[width=6.5cm]{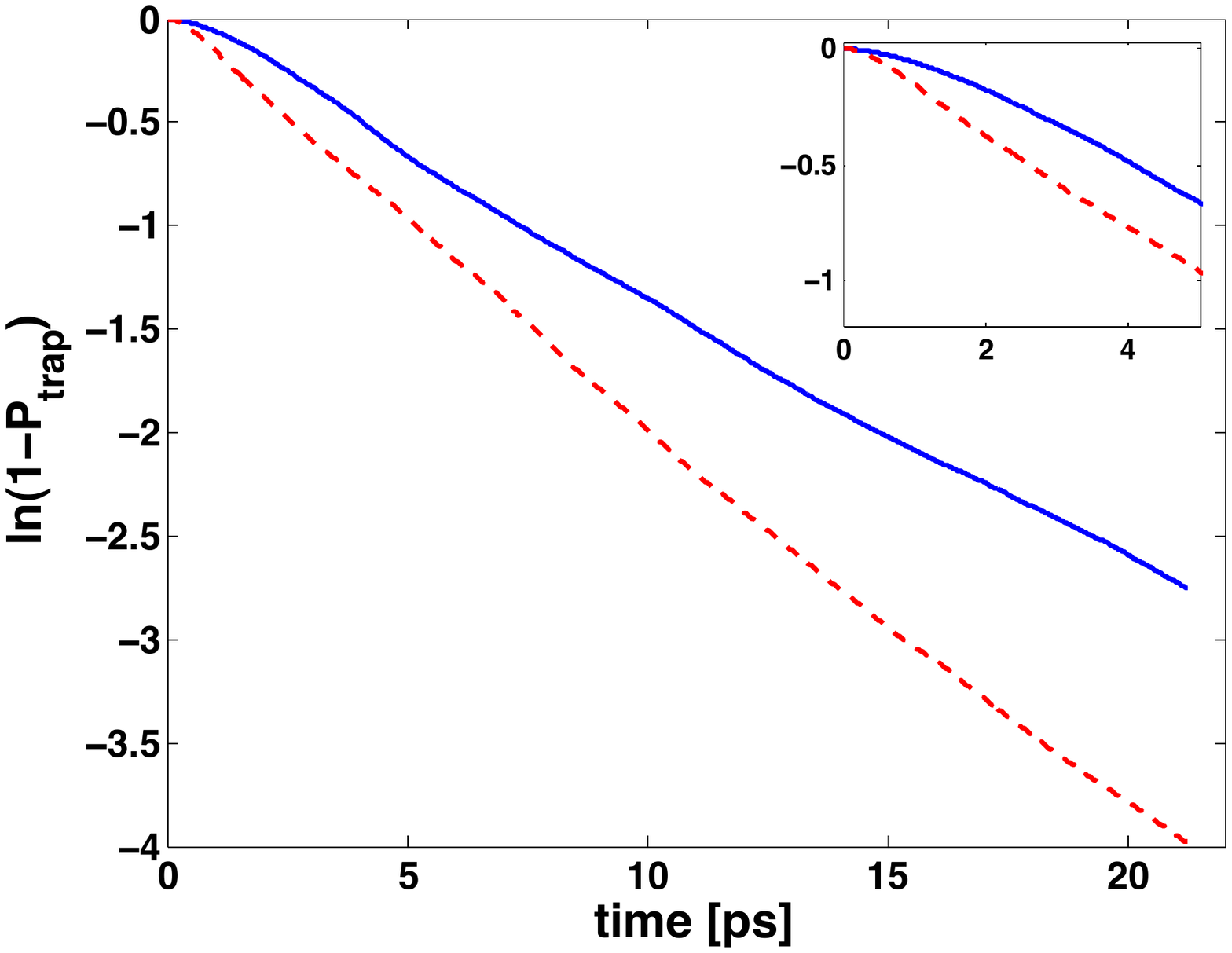}
	\caption{(Color online) Log-linear plot of the average survival probability as a function of time for calculations with environmental correlation time $\tau_c=45$fs and initial state an excitation on BChla 6. This figure only shows the case of no spatial correlations; the temporal scaling is similar for all three models of spatial correlations analyzed in this work. The blue (solid) curve shows behavior at $T=77K$ and the red (dotted) for $T=300K$. The inset is a zoom to the short time region. The average is taken over 100 sample evolutions. Error bars are omitted for clarity: the variation from these average curves is small.}
	\label{fig:ptrap_powerlaw} 
\end{figure}

For completeness,  in Fig. \ref{fig:ptrap} we also show representative time traces of the average trapping probability, for temporal correlation times $\tau_c=45$ fs and $\tau_c=120$fs, and for the three different spatial correlation models. The trapping probability curves have similar trends for all spatiotemporal correlations and the insets show that the dependence on spatial correlation length becomes less pronounced at higher temperatures. Fig. \ref{fig:ptrap_powerlaw} shows the survival probability: $P_{\textrm{surv}} \equiv 1-P_{\textrm{trap}}$ at the two temperatures 77K and 300K, for the case of no spatial correlations and temporal correlation time $\tau_c=45$fs. The time dependence of this survival probability is often examined in the analysis of random walk dynamics with trapping \cite{Mul.Blu.etal-2007}. For the stochastic Liouville description used here and for the graph defined by the FMO complex, we see that at $T=77K$ the survival probability at short times shows a complex modified exponential decay that is well fit (except in the $t\rightarrow 0$ limit) by a function of the form: $P_{\textrm{surv}} \propto e^{\sum_i c_i t^{i}}$ with $i$ ranging from 0 to 3 at least, containing both subexponential ($c_i <0$) and superexponential ($c_i >0$) contributions. The decay at long times fits well to a simple exponential decay $e^{-\beta t}$. At the higher temperature $T=300K$, the modified exponential decay regime is shortened significantly, and most of the decay is simply exponential.

\section{Discussion}
To understand the behavior of trapping efficiency with temporal correlation, we focus on the case of no spatial correlation in fluctuations since the general behavior with respect to $\tau_{c}$ is the same for all spatial correlations considered. We begin by examining the excitonic structure of FMO, shown in Table \ref{tab:excitons}. The wavefunctions of the two lowest energy excitons have significant overlap with the trapping site, BChla 3. Similarly, the initial state of an excitation localized on site 1 (site 6) primarily has components of excitons 2 and 5 (excitons 4 and 6). Using a second-order cumulant expansion technique \footnote{The second order truncation ignores the effects of all third and higher moments of the fluctuating operator in the Hamiltonian.}, Blumen and Silbey have shown that for such a model, when written in the interaction picture with respect to $H_{\textrm{el}}$, the populations in the exciton basis evolve according to the master equation \cite{Blu.Sil-1978}:
\bqa
\frac{\textrm{d}}{\textrm{d}t}\rho_{\alpha\alpha}(t) &=& \frac{1}{N}\sum^{N-1}_{\beta=0} \Gamma_{\alpha\beta}(t) \left( -\rho_{\alpha\alpha}(t) + \rho_{\beta\beta}(t) \right) \nn
\label{eq:master}
\eqa 
where $N$ is the total number of chromophores, and we have used Greek indices for the exciton basis. $\Gamma_{\alpha\beta}$ is a population transition rate from exciton level $\alpha$ to $\beta$, and is explicitly (in our notation):
\bqa
\Gamma_{\alpha\beta}(t) &=& 2\Delta^{2}_{0}\left( \frac{\tau_{c}}{1+\tau_{c}^{2} \omega_{\alpha\beta}^{2}} \right. \nn \\  &+&  \left. \frac{{\tau_c}^{2}e^{-t/\tau_c}}{1+\tau_{c}^{2} \omega_{\alpha\beta}^{2}}\left[ \omega_{\alpha\beta}\sin(\omega_{\alpha\beta}t) - \frac{\cos(\omega_{\alpha\beta}t)}{\tau_{c}}\right] \right) \nn
\label{eq:solution}
\eqa
where $\omega_{\alpha\beta} \equiv \omega_{\alpha}-\omega_{\beta}$ is a difference between exciton frequencies. For times $t \gg \tau_{c}$, which is when most of the excitation transfer occurs, the second, time dependent term in Eq.~(\ref{eq:solution}) is damped by the exponential prefactor and can be neglected. Hence the transition rates are essentially determined by the constant term, which is equal to the Lorentzian power spectrum of the exponentially correlated noise: $J_{\Delta}(\omega) = \frac{2\Delta^{2}_{0}\tau_{c}}{1+\tau_{c}^{2}\omega^{2}}$. In order to maximize transport efficiency, it is advantageous to maximize the rate of transitions between exciton states. The time-independent part of the rates $\Gamma_{\alpha\beta}$ is maximized when the correlation time of the fluctuations matches the energy differences between the exciton levels of the complex -- i.e., $\tau_{c} = 1/\omega_{\alpha\beta}$. Physically, this results from the fact that noise power at an exciton energy difference drives population transitions between the corresponding exciton levels. As can be seen from Table \ref{tab:excitons} the exciton energy differences in FMO are mostly in the region $\Delta E \sim 90 - 350 \textrm{cm}^{-1}$. For this range of energy differences, the range of correlation times that maximizes $J_{\Delta}$ is $\tau_{c} \sim 15 - 60 \textrm{fs}$. This is precisely the range in which the peaks in average efficiency lie in Figs. \ref{fig:i1_77} - \ref{fig:i6_300}. Thus we conclude that the peak in average efficiency with respect to temporal correlation is due to noise assisted transfer. 

\begin{table}[h]
\begin{tabular}{|c||c|c|c|c|c|c|c|}
\hline
Exciton & 0 & 1 & 2 & 3 & 4 & 5 & 6 \\
\hline
Energy & 0 & 102.8 & 177 & 272.7 & 297.5 & 402.7 & 497.2 \\
Site overlap & 3, 4 & 3, 4, 5, 7 & 1, 2 & 5, 7& 4, 5, 6 & 1, 2 & 5, 6 \\
\hline
\end{tabular}
\caption{FMO excitons (energy eigenstates of $H_{\textrm{el}}$). The exciton energies are in $\textrm{cm}^{-1}$ (normalized so that the lowest energy exciton has energy zero). The third row shows the sites that the exciton wavefunction has largest overlap with.}
\label{tab:excitons}
\end{table}

\begin{figure*}[t!]
\centering 
\subfloat[~$T=77$K] { 
\includegraphics[width=7.5cm]{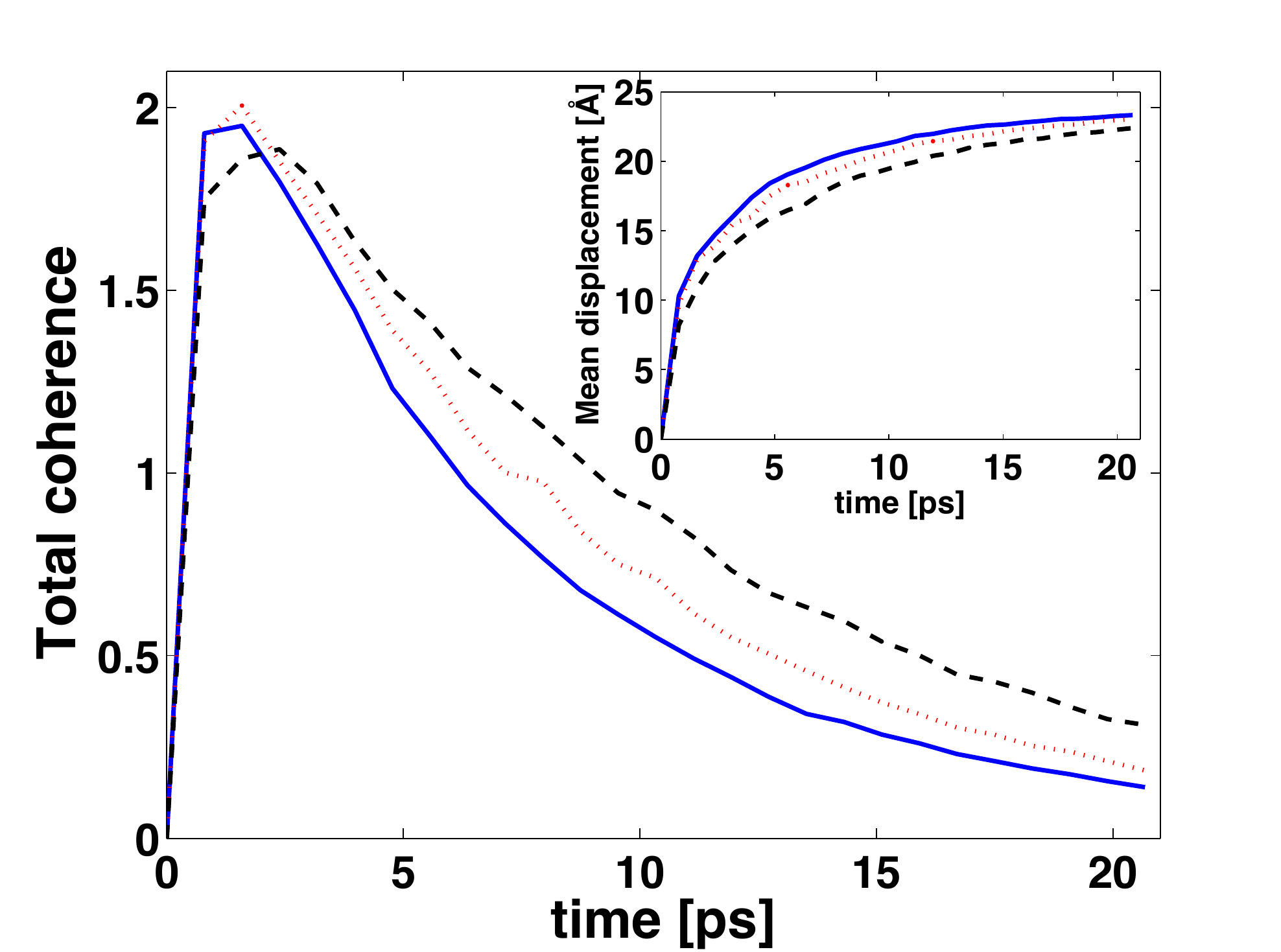}
} 
\hspace{1cm}
\subfloat[~$T=300$K] { 
\includegraphics[width=7.5cm]{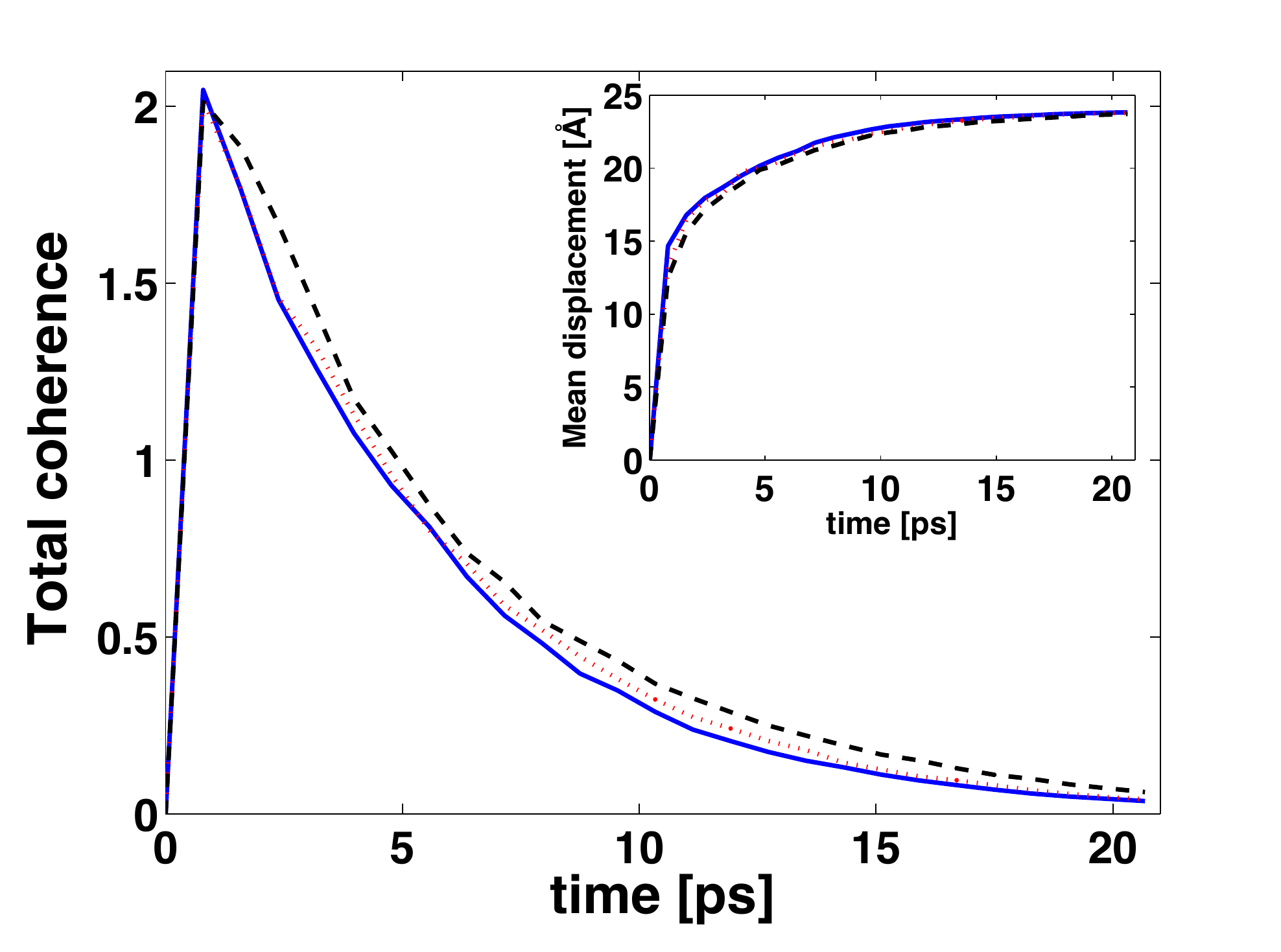}
}
\caption{(Color online) Average total coherence (main axes) and average displacement in $\AA$ (insets) in the FMO complex as a function of time for two temperatures. The initial state is an excitation on BChla 6 and $\tau_{c}=45\textrm{fs}$. The blue (solid) curves are for no spatial correlations, $\mathbf{C}^{N}$; the red (dotted) and black (dashed) curves are for exponential spatial correlations with $R_c=10 \AA$ and $R_c=20\AA$, respectively. The average is taken over 100 sample evolutions. Error bars are omitted for clarity, the variation from these average curves is small. The inset shows the mean displacement $\mathcal{X}$ (defined in the main text) as a function of time for the same spatial correlation cases.} 
\label{fig:spatial}
\end{figure*}

Now we turn to the effects of spatial correlations. The primary mechanism by which energy fluctuations affect excitation transport is by modulating energy mismatches between chromophores \cite{Fle.Cho-1996, Car.Chi.etal-2009}.  The rate of excitation transfer between two chromophores is enhanced when their average energy gap decreases. Positive correlations in the energy fluctuations suppress line broadening and in the limit of perfectly correlated fluctuations the energy gaps remain unchanged. This can equivalently be viewed as a renormalization in which positive correlations reduce the reorganization energy, and hence the dephasing rate \cite{Hen.Bel.etal-2009}. The suppression of line broadening has two key effects on excitations: it leads to longer-lived coherence between sites and it slows the average transport by reducing transfer rates between energy mismatched chromophores. We illustrate both of these effects in Fig. \ref{fig:spatial}. The main axes plot the time evolution of total coherence in FMO, which is defined here as: $\mathcal{C} = \sum_{i\neq j} |\rho_{ij}|$. The initial state is an excitation localized on BChla 6, and the two panels show the decay of coherence at two temperatures.  Both plots clearly show the preservation of coherence by spatially correlated fluctuations, although the effect is less dramatic for the higher temperature. Further, to demonstrate the second point that correlated fluctuations lead to slower excitation transport, in the insets we plot the mean displacement of the excitation, defined as $\mathcal{X}(t) = \sum_{i} d_{6i} \rho_{ii}(t)$, where $d_{6i}$ is the distance between the initial site 6 and site $i$. 

It is clear from these plots that after the first few hundred femtoseconds, the average rate of transfer of the excitation is reduced in the presence of spatially correlated fluctuations. Again, the effect is less dramatic at higher temperatures where the line broadening is inherently larger and the renormalization by spatial correlations has less of an impact. Thus we conclude that positive spatial correlations in environmental noise lead to both longer-lived coherence and slower excitation transport. These conclusions agree with recent studies of the influence of spatial correlations in chromophoric systems using a variety of techniques \cite{Reb.Moh.etal-2009a, Hen.Bel.etal-2009, Fas.Naz.etal-2009, Wu.Liu.etal-2010, Moh.Sha.etal-2010}.

As noted above, in view of the fact that the exact nature of the spatial correlation present in the environmental fluctuations is unknown \cite{Ren.Mar-2002, Ren-2009}, we also simulated excitation dynamics assuming a polynomial instead of exponential dependence of the spatial correlations. In particular, we used $\mathbf{C}^{DB}_{ij} = \frac{1}{d_{ij}^{2}} \times \beta$, i.e., an inverse squared dependency of vibrational correlations in the protein scaffolding on the distance between chromophores. Here $\beta$ is the largest constant $<1$ that ensures that the matrix $\mathbf{C}^{DB}$ is positive definite: the magnitude of $\beta$ thus constitutes an upper bound on the magnitude of correlations between fluctuations on different sites.  With the inter-chromophore distances from Ref. \cite{Cam.Bla.etal-2003}, we find $\beta \approx 0.85$.  Our calculations of the dynamics resulting from this correlation of the environmental fluctuations (not shown here) are very similar to that presented for exponentially decaying correlations in Figures~1-5.  In particular, there is an optimal temporal correlation time of $\sim 40 \textrm{fs}$ for initial state 1 (and $\sim 30 \textrm{fs}$ for initial state 6), and the overall efficiency of energy transfer is seen to be lowered by the spatial correlation, while coherence in the site basis is enhanced, as in Figure~\ref{fig:spatial}.

\section{Conclusion} 
We have systematically studied the effects of temporal and spatial correlations of noise on the transport of excitations across a prototypical light harvesting complex. We emphasize that since we employed a classical stochastic model of the phonon environment, it is not possible to draw quantitative conclusions regarding energy transfer in FMO from the present study. However, our model does enable us to examine the relative effects on transport of varying amounts of correlation. We have shown that temporal correlations can enhance noise power in certain spectral regions and consequently enhance exciton transitions lying in these regions. We also showed that spatially correlated fluctuations can preserve coherence while at the same time resulting in slower transport to the trapping site. While this paper was under review we learned of recent studies using the more physically accurate generalized Bloch-Redfield equation approach \cite{Wu.Liu.etal-2010, Moh.Sha.etal-2010} that arrive at similar conclusions. These authors further investigated the dependence of these phenomena on the reorganization energy of the protein pigment complex and found that depending on the magnitude of the reorganization energy, positive spatial correlations can either decrease or increase the overall efficiency of the energy transfer process.  This interesting result is consistent with both the fact that spatial correlations act to effectively decrease the reorganization energy of the system and the fact that there is an optimal value for the reorganization energy. Thus depending whether the reorganization energy is smaller or larger than the optimal value, spatial correlations may shift the value either towards the optimal, increasing the efficiency, or away from it, decreasing the efficiency. Finally, although we have specifically analyzed the FMO complex in this work, the conclusions drawn here about the effects of correlated fluctuations on excitation transport will also apply to larger light harvesting complexes and are generally applicable to transport phenomena in densely packed molecular assemblies, including J-aggregates and other photosynthetic units \cite{Kob-1996}.

This study also raises several intriguing questions of biological import. Most significantly, why is it that spatially correlated fluctuations are likely present in some LHCs \cite{Eng.Cal.etal-2007, Lee.Che.etal-2007} if they serve to {\em reduce} rather than to enhance the efficiency and speed of excitation transport? Second, is the correlation time of environmental fluctuations in FMO within the window that maximizes transport efficiency? Clearly, more detailed studies are warranted of the complex environments for electronic energy transport that are found in natural organic molecular assemblies, in particular for light harvesting complexes.

\section*{Acknowledgements}
We would like to acknowledge useful discussions with Akihito Ishizaki and Robert Silbey on the subject of correlated fluctuations and energy transfer. This material is based upon work supported by DARPA under Award No. N66001-09-1-2026.


\begin{thebibliography}{32}
\expandafter\ifx\csname natexlab\endcsname\relax\def\natexlab#1{#1}\fi
\expandafter\ifx\csname bibnamefont\endcsname\relax
  \def\bibnamefont#1{#1}\fi
\expandafter\ifx\csname bibfnamefont\endcsname\relax
  \def\bibfnamefont#1{#1}\fi
\expandafter\ifx\csname citenamefont\endcsname\relax
  \def\citenamefont#1{#1}\fi
\expandafter\ifx\csname url\endcsname\relax
  \def\url#1{\texttt{#1}}\fi
\expandafter\ifx\csname urlprefix\endcsname\relax\def\urlprefix{URL }\fi
\providecommand{\bibinfo}[2]{#2}
\providecommand{\eprint}[2][]{\url{#2}}

\bibitem[{\citenamefont{Engel et~al.}(2007)\citenamefont{Engel, Calhoun, Read,
  Ahn, Mancal, Cheng, Blankenship, and Fleming}}]{Eng.Cal.etal-2007}
\bibinfo{author}{\bibfnamefont{G.~S.} \bibnamefont{Engel}}
 \textit{et~al.}, \bibinfo{journal}{Nature}
  \textbf{\bibinfo{volume}{446}}, \bibinfo{pages}{782} (\bibinfo{year}{2007}).

\bibitem[{\citenamefont{Lee et~al.}(2007)\citenamefont{Lee, Cheng, and
  Fleming}}]{Lee.Che.etal-2007}
\bibinfo{author}{\bibfnamefont{H.}~\bibnamefont{Lee}},
  \bibinfo{author}{\bibfnamefont{Y.-C.} \bibnamefont{Cheng}},
  \bibinfo{author}{\bibfnamefont{G.~R.} \bibnamefont{Fleming}},
  \bibinfo{journal}{Science} \textbf{\bibinfo{volume}{316}},
  \bibinfo{pages}{1462} (\bibinfo{year}{2007}).
  
\bibitem{Col.Won.etal-2010}
E. Collini, \textit{et al.},
\newblock{Nature}, \textbf{463}, 644 (2010).

\bibitem{Pan.Hay.etal-2010}
G. Panitchayangkoon, \textit{et al.},
\newblock{Proc. Natl. Acad. Sci. USA}, \textbf{107}, 12766 (2010).

\bibitem[{\citenamefont{Collini and Scholes}(2009)}]{Col.Sch-2009}
\bibinfo{author}{\bibfnamefont{E.}~\bibnamefont{Collini}},
  \bibinfo{author}{\bibfnamefont{G.~D.} \bibnamefont{Scholes}},
  \bibinfo{journal}{Science} \textbf{\bibinfo{volume}{323}},
  \bibinfo{pages}{369} (\bibinfo{year}{2009}).

\bibitem[{\citenamefont{Gaab and Bardeen}(2004)}]{Gaa.Bar-2004}
\bibinfo{author}{\bibfnamefont{K.~M.} \bibnamefont{Gaab}},
  \bibinfo{author}{\bibfnamefont{C.~J.} \bibnamefont{Bardeen}},
  \bibinfo{journal}{J. Chem. Phys.} \textbf{\bibinfo{volume}{121}},
  \bibinfo{pages}{7813} (\bibinfo{year}{2004}).

\bibitem[{\citenamefont{Rebentrost
  et~al.}(2009{\natexlab{a}})\citenamefont{Rebentrost, Mohseni, Kassal, Lloyd,
  and Aspuru-Guzik}}]{Reb.Moh.etal-2009}
\bibinfo{author}{\bibfnamefont{P.}~\bibnamefont{Rebentrost}}
\textit{et~al.},
  \bibinfo{journal}{New J. Phys.} \textbf{\bibinfo{volume}{11}},
  \bibinfo{pages}{033003} (\bibinfo{year}{2009}{\natexlab{a}}).

\bibitem[{\citenamefont{Mohseni et~al.}(2008)\citenamefont{Mohseni, Rebentrost,
  Lloyd, and Aspuru-Guzik}}]{Moh.Reb.etal-2008}
\bibinfo{author}{\bibfnamefont{M.}~\bibnamefont{Mohseni}}
\textit{et~al.},
  \bibinfo{journal}{J. Chem. Phys.} \textbf{\bibinfo{volume}{129}},
  \bibinfo{pages}{174106} (\bibinfo{year}{2008}).

\bibitem[{\citenamefont{Rebentrost
  et~al.}(2009{\natexlab{b}})\citenamefont{Rebentrost, Mohseni, and
  Aspuru-Guzik}}]{Reb.Moh.etal-2009a}
\bibinfo{author}{\bibfnamefont{P.}~\bibnamefont{Rebentrost}},
  \bibinfo{author}{\bibfnamefont{M.}~\bibnamefont{Mohseni}},
  \bibinfo{author}{\bibfnamefont{A.}~\bibnamefont{Aspuru-Guzik}},
  \bibinfo{journal}{J. Phys. Chem. B} \textbf{\bibinfo{volume}{113}},
  \bibinfo{pages}{9942} (\bibinfo{year}{2009}{\natexlab{b}}).
  
\bibitem[{\citenamefont{Plenio and Huelga}(2008)}]{Ple.Hue-2008}
\bibinfo{author}{\bibfnamefont{M.~B.} \bibnamefont{Plenio}},
  \bibinfo{author}{\bibfnamefont{S.~F.} \bibnamefont{Huelga}},
  \bibinfo{journal}{New J. Phys.} \textbf{\bibinfo{volume}{10}},
  \bibinfo{pages}{113019} (\bibinfo{year}{2008}).

\bibitem[{\citenamefont{Caruso et~al.}(2009)\citenamefont{Caruso, Chin, Datta,
  Huelga, and Plenio}}]{Car.Chi.etal-2009}
\bibinfo{author}{\bibfnamefont{F.}~\bibnamefont{Caruso}}
\textit{et~al.},
  \bibinfo{journal}{J. Chem. Phys.} \textbf{\bibinfo{volume}{131}},
  \bibinfo{pages}{105106} (\bibinfo{year}{2009}).

\bibitem[{\citenamefont{Ishizaki and
  Fleming}(2009{\natexlab{a}})}]{Ish.Fle-2009a}
\bibinfo{author}{\bibfnamefont{A.}~\bibnamefont{Ishizaki}},
  \bibinfo{author}{\bibfnamefont{G.~R.} \bibnamefont{Fleming}},
  \bibinfo{journal}{J. Chem. Phys.} \textbf{\bibinfo{volume}{130}},
  \bibinfo{pages}{234110} (\bibinfo{year}{2009}{\natexlab{a}}).

\bibitem[{\citenamefont{Renger et~al.}(2001)\citenamefont{Renger, May, and
  Kuhn}}]{Ren.May.etal-2001}
\bibinfo{author}{\bibfnamefont{T.}~\bibnamefont{Renger}},
  \bibinfo{author}{\bibfnamefont{V.}~\bibnamefont{May}},
  \bibinfo{author}{\bibfnamefont{O.}~\bibnamefont{Kuhn}},
  \bibinfo{journal}{Phys. Rep.} \textbf{\bibinfo{volume}{343}},
  \bibinfo{pages}{138} (\bibinfo{year}{2001}).

\bibitem[{\citenamefont{Scholes and Fleming}(2006)}]{Sch.Fle-2005}
\bibinfo{author}{\bibfnamefont{G.~D.} \bibnamefont{Scholes}},
  \bibinfo{author}{\bibfnamefont{G.~R.} \bibnamefont{Fleming}},
  \bibinfo{journal}{Adv. Chem. Phys.} \textbf{\bibinfo{volume}{132}},
  \bibinfo{pages}{57} (\bibinfo{year}{2005}).

\bibitem[{\citenamefont{Cheng and Fleming}(2009)}]{Che.Fle-2009}
\bibinfo{author}{\bibfnamefont{Y.-C.} \bibnamefont{Cheng}},
  \bibinfo{author}{\bibfnamefont{G.~R.} \bibnamefont{Fleming}},
  \bibinfo{journal}{Ann. Rev. Phys. Chem.} \textbf{\bibinfo{volume}{60}},
  \bibinfo{pages}{241} (\bibinfo{year}{2009}).

\bibitem[{\citenamefont{Cogdell et~al.}(2006)\citenamefont{Cogdell, Gall, and
  K{\"o}hler}}]{Cog.Gal.etal-2006}
\bibinfo{author}{\bibfnamefont{R.~J.} \bibnamefont{Cogdell}},
  \bibinfo{author}{\bibfnamefont{A.}~\bibnamefont{Gall}},
  \bibinfo{author}{\bibfnamefont{J.}~\bibnamefont{K{\"o}hler}},
  \bibinfo{journal}{Q. Rev. Biophys.} \textbf{\bibinfo{volume}{39}},
  \bibinfo{pages}{227} (\bibinfo{year}{2006}).

\bibitem[{\citenamefont{Muh et~al.}(2007)\citenamefont{Muh, Madjet, Adolphs,
  Abdurahman, Rabenstein, Ishikita, Knapp, and Renger}}]{Muh.Mad.etal-2007}
\bibinfo{author}{\bibfnamefont{F.}~\bibnamefont{Muh}}
\textit{et~al.},
  \bibinfo{journal}{Proc. Nat. Acad. Sc.} \textbf{\bibinfo{volume}{104}},
  \bibinfo{pages}{16862} (\bibinfo{year}{2007}).

\bibitem[{\citenamefont{Renger and Marcus}(2002)}]{Ren.Mar-2002}
\bibinfo{author}{\bibfnamefont{T.}~\bibnamefont{Renger}},
  \bibinfo{author}{\bibfnamefont{R.~A.} \bibnamefont{Marcus}},
  \bibinfo{journal}{J. Chem. Phys.} \textbf{\bibinfo{volume}{116}},
  \bibinfo{pages}{9997} (\bibinfo{year}{2002}).

\bibitem[{\citenamefont{Fleming and Cho}(1996)}]{Fle.Cho-1996}
\bibinfo{author}{\bibfnamefont{G.~R.} \bibnamefont{Fleming}},
  \bibinfo{author}{\bibfnamefont{M.}~\bibnamefont{Cho}},
  \bibinfo{journal}{Annu. Rev. Phys. Chem.} \textbf{\bibinfo{volume}{47}},
  \bibinfo{pages}{109} (\bibinfo{year}{1996}).

\bibitem[{\citenamefont{Ishizaki and
  Fleming}(2009{\natexlab{b}})}]{Ish.Fle-2009}
\bibinfo{author}{\bibfnamefont{A.}~\bibnamefont{Ishizaki}},
  \bibinfo{author}{\bibfnamefont{G.~R.} \bibnamefont{Fleming}},
  \bibinfo{journal}{J. Chem. Phys.} \textbf{\bibinfo{volume}{130}},
  \bibinfo{pages}{234111} (\bibinfo{year}{2009}{\natexlab{b}}).

\bibitem[{\citenamefont{Nazir}(2009)}]{Naz-2009}
\bibinfo{author}{\bibfnamefont{A.}~\bibnamefont{Nazir}},
  \bibinfo{journal}{Phys. Rev. Lett.} \textbf{\bibinfo{volume}{103}},
  \bibinfo{pages}{146404} (\bibinfo{year}{2009}).
  
\bibitem[{\citenamefont{Fenna and Matthews}(1975)}]{Fen.Mat-1975}
\bibinfo{author}{\bibfnamefont{R.~E.} \bibnamefont{Fenna}},
  \bibinfo{author}{\bibfnamefont{B.~W.} \bibnamefont{Matthews}},
  \bibinfo{journal}{Nature} \textbf{\bibinfo{volume}{258}},
  \bibinfo{pages}{573} (\bibinfo{year}{1975}).

\bibitem[{\citenamefont{Camara-Artigas
  et~al.}(2003)\citenamefont{Camara-Artigas, Blankenship, and
  Allen}}]{Cam.Bla.etal-2003}
\bibinfo{author}{\bibfnamefont{A.}~\bibnamefont{Camara-Artigas}},
  \bibinfo{author}{\bibfnamefont{R.~E.} \bibnamefont{Blankenship}},
  \bibinfo{author}{\bibfnamefont{J.~P.} \bibnamefont{Allen}},
  \bibinfo{journal}{Photosynth. Res.} \textbf{\bibinfo{volume}{75}},
  \bibinfo{pages}{49} (\bibinfo{year}{2003}).

\bibitem[{\citenamefont{Li et~al.}(1997)\citenamefont{Li, Zhou, Blankenship,
  and Allen}}]{Li.Zho.etal-1997}
\bibinfo{author}{\bibfnamefont{Y.~F.} \bibnamefont{Li}}
\textit{et~al.},
  \bibinfo{journal}{J. Mol. Biol.} \textbf{\bibinfo{volume}{271}},
  \bibinfo{pages}{456} (\bibinfo{year}{1997}).

\bibitem[{\citenamefont{Adolphs and Renger}(2006)}]{Ado.Ren-2006}
\bibinfo{author}{\bibfnamefont{J.}~\bibnamefont{Adolphs}},
  \bibinfo{author}{\bibfnamefont{T.}~\bibnamefont{Renger}},
  \bibinfo{journal}{Biophysical J.} \textbf{\bibinfo{volume}{91}},
  \bibinfo{pages}{2778} (\bibinfo{year}{2006}).

\bibitem[{\citenamefont{Wen et~al.}(2009)\citenamefont{Wen, Zhang, Gross, and
  Blankenship}}]{Wen.Zha.etal-2009}
\bibinfo{author}{\bibfnamefont{J.}~\bibnamefont{Wen}}
\textit{et~al.},
  \bibinfo{journal}{Proc. Nat. Acad. Sc.} \textbf{\bibinfo{volume}{106}},
  \bibinfo{pages}{6134} (\bibinfo{year}{2009}).

\bibitem[{\citenamefont{Haken and Strobl}(1973)}]{Hak.Str-1973}
\bibinfo{author}{\bibfnamefont{H.}~\bibnamefont{Haken}},
  \bibinfo{author}{\bibfnamefont{G.}~\bibnamefont{Strobl}},
  \bibinfo{journal}{Z. Phys.} \textbf{\bibinfo{volume}{262}},
  \bibinfo{pages}{135} (\bibinfo{year}{1973}).

\bibitem[{\citenamefont{Silbey}(1976)}]{Sil-1976}
\bibinfo{author}{\bibfnamefont{R.}~\bibnamefont{Silbey}},
  \bibinfo{journal}{Ann. Rev. Phys. Chem.} \textbf{\bibinfo{volume}{27}},
  \bibinfo{pages}{203} (\bibinfo{year}{1976}).

\bibitem[{\citenamefont{Rips}(1993)}]{Rip-1993}
\bibinfo{author}{\bibfnamefont{I.}~\bibnamefont{Rips}}, \bibinfo{journal}{Phys.
  Rev. E} \textbf{\bibinfo{volume}{47}}, \bibinfo{pages}{67}
  (\bibinfo{year}{1993}).

\bibitem[{\citenamefont{Mukamel}(1999)}]{Muk-1999}
\bibinfo{author}{\bibfnamefont{S.}~\bibnamefont{Mukamel}},
  \emph{\bibinfo{title}{Principles of nonlinear optical spectroscopy}}
  (\bibinfo{publisher}{Oxford University Press}, \bibinfo{year}{1999}).

\bibitem[{\citenamefont{Brixner et~al.}(2005)\citenamefont{Brixner, Stenger,
  Vaswani, Cho, Blankenship, and Fleming}}]{Bri.Ste.etal-2005}
\bibinfo{author}{\bibfnamefont{T.}~\bibnamefont{Brixner}}
\textit{et~al.}, \bibinfo{journal}{Nature}
  \textbf{\bibinfo{volume}{434}}, \bibinfo{pages}{625} (\bibinfo{year}{2005}).

\bibitem[{\citenamefont{Cho et~al.}(2005)\citenamefont{Cho, Vaswani, Brixner,
  Stegner, and Fleming}}]{Cho.Vas.etal-2005}
\bibinfo{author}{\bibfnamefont{M.}~\bibnamefont{Cho}}
\textit{et~al.},
  \bibinfo{journal}{J. Phys. Chem. B} \textbf{\bibinfo{volume}{109}},
  \bibinfo{pages}{10542} (\bibinfo{year}{2005}).

\bibitem[{\citenamefont{Blumen and Silbey}(1978)}]{Blu.Sil-1978}
\bibinfo{author}{\bibfnamefont{A.}~\bibnamefont{Blumen}},
  \bibinfo{author}{\bibfnamefont{R.}~\bibnamefont{Silbey}},
  \bibinfo{journal}{J. Chem. Phys.} \textbf{\bibinfo{volume}{69}},
  \bibinfo{pages}{3589} (\bibinfo{year}{1978}).

\bibitem[{\citenamefont{Hennebicq et~al.}(2009)\citenamefont{Hennebicq,
  Beljonne, Curutchet, Scholes, and Silbey}}]{Hen.Bel.etal-2009}
\bibinfo{author}{\bibfnamefont{E.}~\bibnamefont{Hennebicq}}
\textit{et~al.}, \bibinfo{journal}{J. Chem. Phys.}
  \textbf{\bibinfo{volume}{130}}, \bibinfo{pages}{214505}
  (\bibinfo{year}{2009}).

\bibitem[{\citenamefont{Nitzan}(2006)}]{Nit-2006}
\bibinfo{author}{\bibfnamefont{A.}~\bibnamefont{Nitzan}},
  \emph{\bibinfo{title}{Chemical dynamics in condensed phases }}
  (\bibinfo{publisher}{Oxford University Press}, \bibinfo{year}{2006}).

\bibitem[{\citenamefont{Fassiolo et~al.}(2009)\citenamefont{Fassiolo, Nazir,
  and Olaya-Castro}}]{Fas.Naz.etal-2009}
\bibinfo{author}{\bibfnamefont{F.}~\bibnamefont{Fassiolo}},
  \bibinfo{author}{\bibfnamefont{A.}~\bibnamefont{Nazir}}, \bibnamefont{and}
  \bibinfo{author}{\bibfnamefont{A.}~\bibnamefont{Olaya-Castro}}
  (\bibinfo{year}{2009}), \eprint{arXiv:0907.5183 [quant-ph]}.
  
\bibitem{Ren.May-1998}
T.~Renger and V.~May.
\newblock {J. Phys. Chem. A}, \textbf{102}, 4381 (1998).

\bibitem{Wu.Liu.etal-2010}
J.~Wu, F.~Liu, Y.~Shen, J.~Cao, and R.~J. Silbey.
\newblock{New. J. Phys.}, \textbf{12}, 105012 (2010).
\newblock arXiv:1008.2236 [physics.chem-ph].

\bibitem{Moh.Sha.etal-2010}
M.~Mohseni, A.~Shabani, S.~Lloyd and H.~Rabitz.
\newblock (Manuscript in preparation) (2010).
 
\bibitem{Ren-2009}
T.~Renger.
\newblock Photosynth. Res., \textbf{102}, 471 (2009).

\bibitem{Ish.Fle-2009b}
A.~Ishizaki and G.~R.~Fleming.
\newblock{Proc. Natl. Acad. Sci. USA}, \textbf{106}, 17255 (2009).
  
\bibitem{vanK-2007}
\bibinfo{author}{\bibfnamefont{N.~G.}~\bibnamefont{van Kampen}},
  \emph{\bibinfo{title}{Stochastic processes in physics and chemistry}}
  (\bibinfo{publisher}{Elsevier}, \bibinfo{year}{2007}). 
  
\bibitem{Bel.Cur.etal-2009}
D. Beljonne, C. Curutchet, G. D. Scholes, R. J. Silbey.
\newblock{J. Phys. Chem. B}, \textbf{113}, 6583 (2009).

\bibitem{Des-2002}
M. Deserno. \textit{How to generate exponentially correlated Gaussian random numbers}.
http://www.cmu.edu/biolphys/deserno/pdf/\\corr\_gaussian\_random.pdf
  
\bibitem{Joh-1994}
G. E. Johnson. Proc. IEEE, \textbf{82}, 270 (1994). 

\bibitem{Mul.Blu.etal-2007}
O. Mulken, A. Blumen, T. Amthor, C. Giese, M. Reetz-Lamour, M. Weidemuller.
\newblock{Phys. Rev. Lett.}, \textbf{99}, 090601 (2007).

\bibitem[{\citenamefont{Kobayashi}(1996)}]{Kob-1996}
  \emph{\bibinfo{title}{J-aggregates}}.
\bibinfo{editor}{\bibfnamefont{T.}~\bibnamefont{Kobayashi} (Ed.)},
  (\bibinfo{publisher}{World Scientific}, \bibinfo{year}{1996}).
  
\end{thebibliography}
\end{document}